\documentclass[a4paper,11pt]{article}
\pdfoutput=1 

\usepackage{jinstpub} 
\usepackage{textgreek}
\usepackage{caption}
\usepackage{subcaption}
\usepackage{color}
\usepackage{hyperref}
\usepackage[utf8]{inputenc}
\usepackage[acronym]{glossaries}
\glsdisablehyper
\usepackage{enumitem}

\usepackage{textcomp} 

\setlist{itemsep=0pt,parsep=0pt}

\makeglossaries
\newacronym{clb}{CLB}{Central Logic Board}
\newacronym{fpga}{FPGA}{Field Programmable Gate Array}
\newacronym{sfp}{SFP}{Small Form-factor Pluggable}
\newacronym{dom}{DOM}{Digital Optical Module}
\newacronym{hdl}{HDL}{Hardware Description Language}
\newacronym{pmt}{PMT}{Photo Multiplier Tube}
\newacronym{tdc}{TDC}{Time to Digital Converter}
\newacronym{aes}{AES}{Audio Engineering Society}
\newacronym{ptp}{PTP}{Precision Timing Protocol}
\newacronym{lm32}{LM32}{LatticeMico32}
\newacronym{srp}{SRP}{Simple Retransmission Protocol}
\newacronym{mcf}{MCF}{Message Container Format}
\newacronym{daq}{DAQ}{Data AcQuisition}
\newacronym{mac}{MAC}{Media Access Control}
\newacronym{udp}{UDP}{User Datagram Protocol}
\newacronym{dhcp}{DHCP}{Dynamic Host Configuration Protocol}
\newacronym{icmp}{ICMP}{Internet Control Message Protocol}
\newacronym{arp}{ARP}{Address Resolution Protocol}

\title{\boldmath Embedded software developments in KM3NeT phase I}

\author[a,1]{V.~van~Beveren,\note{Corresponding author.}}
\author[b]{D.~Real}
\author[c]{T.~Chiarusi}
\author[b]{D.~Calvo}
\author[d]{S.~Mastroianni}
\author[e]{P.~Musico}
\author[c]{G.~Pellegrini}
\author[a]{P.~Jansweijer}
\author[f]{S.~Colonges}
\author[g]{C.~Bozza}
\author[c]{F.~Filippini}
\author[h]{C.~Nicolau}
\author[i]{A.~Díaz}

\affiliation[a]{Nikhef, National Institute for Subatomic Physics, PO Box 41882, Amsterdam, 1009 DB Netherlands}
\affiliation[b]{IFIC - Instituto de F{\'\i}sica Corpuscular (CSIC - Universitat de Val{\`e}ncia), c/Catedr{\'a}tico Jos{\'e} Beltr{\'a}n, 2, 46980 Paterna, Valencia, Spain}
\affiliation[c]{INFN, Sezione di Bologna, v.le C. Berti-Pichat, 6/2, Bologna, 40127 Italy}
\affiliation[d]{INFN, Sezione di Napoli, Complesso Universitario di Monte S. Angelo, Via Cintia ed. G,
	Napoli, 80126 Italy}
\affiliation[e]{INFN, Sezione di Genova, Via Dodecaneso 33, Genova, 16146 Italy}
\affiliation[f]{APC, Universit{\'e} Paris Diderot, CNRS/IN2P3, CEA/IRFU, Observatoire de Paris, Sorbonne Paris Cit\'e, 75205 Paris, France} 
\affiliation[g]{Universit{\`a} di Salerno e INFN Gruppo Collegato di Salerno, Dipartimento di Fisica, Via Giovanni Paolo II 132, Fisciano, 84084 Italy}
\affiliation[h]{INFN, Sezione di Roma, Piazzale Aldo Moro 2, Roma, 00185 Italy}
\affiliation[i]{INFN, Sezione di Catania, Via Santa Sofia 64, Catania, 95123 Italy}

\emailAdd{v.van.beveren@nikhef.nl}

\abstract{The KM3NeT Collaboration has already produced more than one thousand acquisition boards, used for building two deep-sea neutrino detectors at the bottom of the Mediterranean Sea, with the aim of instrumenting a volume of several cubic kilometers with light sensors to detect the Cherenkov radiation produced in neutrino interactions. The the so-called Digital Optical Modules, house the PMTs and the acquisition and control electronics of the module, the Central Logic Board, which  includes a Xilinx FPGA and embedded soft processor. The present work presents the architecture and functionalities of the software embedded in the soft processor of the Central Logic Board.}


\collaboration[c]{on behalf of the KM3NeT collaboration}

\proceeding{Very Large Volume Neutrino Telescope Conference\\
  18-21 of May 2021\\
  Valencia (Online)}

\begin{document}
\maketitle
\flushbottom

\printglossary[type=\acronymtype]

\section{Embedded software in the KM3NeT detector}
\label{sec:sw_km3net}
The KM3NeT detector currently being constructed in the Mediterranean Sea\cite{Adri_n_Mart_nez_2016}, consists of various modules which need to be actively controlled. For this purpose the \gls{clb} and its firmware has been designed during the first phase (Phase I\cite{km3net_strategy}) of detector construction and deployment. The \gls{clb} is an electronics board with a Xilinx Kintex 7 \gls{fpga} as central controller. Additionally, it contains an optical transceiver for network connectivity and, various sensors and connectors for attaching data acquisition hardware and instrumentation. The design of the CLB is primarily driven by the need to fit into a KM3NeT \gls{dom}, but it is not specific to it. The CLB is also used as the main controller in Detector Unit Base (DU-Base), and the Calibration Unit Base (CU-Base). Depending in which module the CLB is located different firmware can be loaded, specific to its purpose. 

The \gls{fpga} firmware contains all digital logic required to perform its function and is a composition of gateware and embedded software. Gateware is the logic present inside the programmable logic of the \gls{fpga} fabric and generally written in a hardware description language, while the embedded software is written in a programming language and running on a processor. %
The gateware contains three DAQ units: The \gls{tdc} receiving pulses from the \glspl{pmt} and conditioning hardware, the \gls{aes}-standard encoder receiving data from the acoustic sensor (a hydrophone), and finally MONitoring, which generates run-time performance metrics each time-slice, a $10\,$ms interval. The \gls{daq} modules stream data into the HWStateMachine, which subsequently annotates and frames the data with each time-slice and finally forwards the data from each DAQ module to the IPMux in using round-robin scheduler. The IPMux wraps the annotated data into \gls{udp} packets and dispatches them through the WhiteRabbit \gls{ptp}-core \gls{mac} over the seafloor network to the shore data processing facility. To prevent loss of data buffers are placed in the data-path which fit even the maximum possible data-rates. In addition to \gls{mac}-functions, the WhiteRabbit core is responsible for providing a detector-wide synchronized clock\cite{whiterabbit2011}. 

The gateware contains two \gls{lm32} micro-processors\cite{lm32}, coded in \gls{hdl}: the WhiteRabbit processor for timing control, and the CLB processor for DAQ control and sensor readout. The IPMux acts as a gateway to the detector network for the CLB LM32 software, allowing both reception and transmission of Ethernet packets. This channel is used for the slow-control and for network functions such as \gls{arp}, \gls{dhcp} and \gls{icmp}. This work is mostly focused on the CLB LM32 processor, which runs the KM3NeT specific software. For more information about electronics and readout, the reader is referred to \cite{detcon19}. 

The KM3NeT telescope is a heterogeneous detector and there is not one firmware which can run on all its modules. The firmware comprises of three types of binaries:
\begin{itemize}
	\item The hardware specific FPGA bit-file, CLB v2 or CLB v4\footnote{CLB v4 is a newer version of the CLB v2, containing some different peripheral components, due to either deprecation or feature request.}.
	\item The WhiteRabbit embedded software for either a KM3NeT custom version (denoted as `broadcast') \cite{pellegrino2016}, or standard WhiteRabbit.
	\item The CLB application software which can be either DOM, DU-Base, CU-Base or Golden, for start-up. 
\end{itemize}
When building the firmware, different combinations of such three-types of binaries are merged into a single FPGA firmware image, creating multiple firmware products, each specific to an application and hardware version. The CLB can store up to four different firmware images, but only two are needed: The golden image and the run-time image. Though most firmware images are for a specific module of the detector, the golden image is generic to all modules. It is the first firmware image loaded after powering the detector and contains only minimal hardware initialization. Its purpose is to boot the actual run-time application, but also to allow firmware updates in case the application requires updates, or is not operational. The run-time image is specific to the module the CLB is contained in. Firmware update currently is a manual procedure using an specific firmware update application. The application checks compatibility before programming the updated image.

The First Generation (FG) of the KM3NeT firmware has been in development since 2012 and was deployed in the beginning of 2016. Since 2019 development of the Next Generation (NG) Firmware has started, containing many improvements with respect to the first generation, both in project semantics and program architecture. Project improvements include usage of modern development methodologies and tools such as the usage GIT with sub-modules, Continuous Integration and Docker containers, for a consistent and reusable build environment. Program architecture improvements include a refined software state-machine, integration with of the current production version of WhiteRabbit, and improved error handling. Some of these features which will be covered in more detail ahead in the text.

\section{Software Architecture}
The KM3NeT embedded software is a bare metal application, using no preemptive operating system due to the already limited RAM resources and absence of a threading requirement. When compiled the total program, including runtime memory, takes up less than $256\,$KiB, limited by available RAM blocks on the FPGA. The application is mostly written in C with some assembly for start-up and interrupt handling. The software is split into two main layers: the system-software layer and the application layer. For each kind of firmware image the system-software layer is the same. This layer contains OS-like features, such as a simple cooperative multitasking scheduler, a firmware update unit, various peripheral drivers, a UDP based network stack and support utilities for logging and error handling. The application layer contains a software state-machine and a number of subsystems, each responsible for a different aspect of the application. 

\begin{figure}
	\centering
	\includegraphics[scale=1]{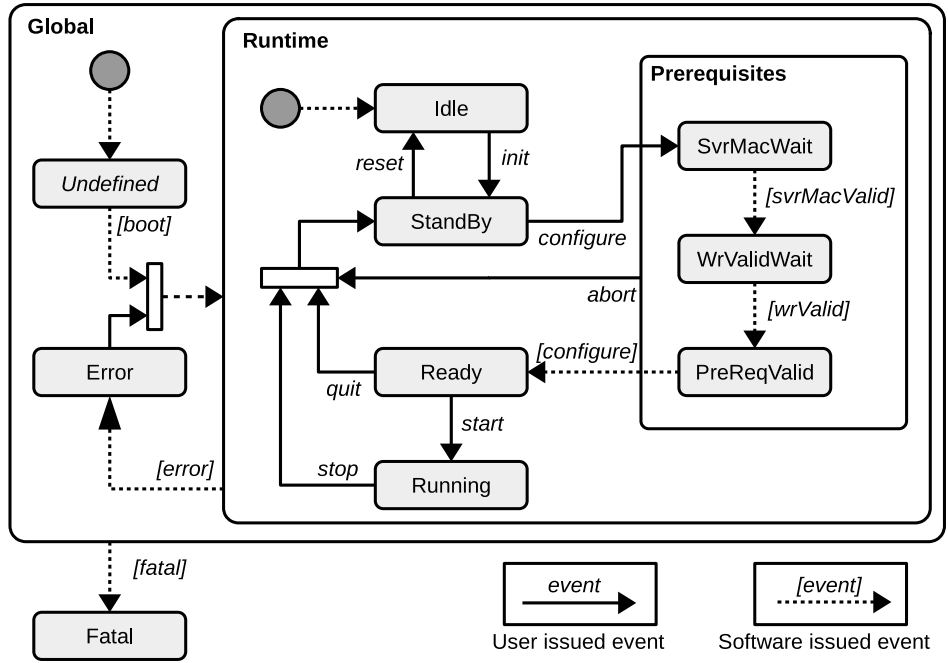}
	\caption{The Next Generation Software State Machine}
	\label{fig:sw_state_machine}
\end{figure}

The state-machine, shown in Figure~\ref{fig:sw_state_machine}, drives state of the application and outside of the control of state-machine little application-level code is run. Implementing the application software as a state-machine obligates the program to always be in a clear and consistent state. The state-machine can be moved by issuing events, generally driven over slow-control by the Control Unit\cite{cu20} application running on shore. Some events are issued autonomously by the embedded software, such as when an error occurs or during system start-up. The application code can be attached to state-machine transitions or to a timer, which calls the code periodically. In case of software stalling a hardware watchdog will reboot the system after $60\,$s of inactivity.

As previously mentioned, application code is grouped into \emph{subsystems}. A subsystem is a unit of code and data responsible for a specific aspect of CLB operation. It controls hardware peripherals, through a driver abstraction, associated with that aspect. The following subsystems are known to the CLB software:
\begin{itemize}
	\item \textbf{System} – Application function not specific to other subsystems
	\item \textbf{Optics} – Control of \glspl{pmt} and \glspl{tdc} (only in DOM firmware)
	\item \textbf{Acoustics} – Acoustic sensor and AES control
	\item \textbf{Instrumentation} – Sensor readout, generally over the i2c bus (Temperature, humidity, etc.)
	\item \textbf{Networking} – IPMux control and WhiteRabbit monitoring
	\item \textbf{Base} – DU-Base container control (only in DU-Base firmware).
\end{itemize}
 By registering C-functions to state transitions a subsystem can control hardware at specific points in the state-machine graph. For example, the \textit{start} event moves the state-machine from Ready to Running. In this transition the data-acquisition hardware is enabled to start data taking.

The CLB slow control from remote is implemented by means of a custom  protocol, on top of UDP. The Slow-control protocol consists of three layers. The highest layer is called the Message layer and binds to C-functions at the application level. Messages have a type, e.g. `retrieve firmware version' or `state-machine event', but also a class, being either Command, Reply, Event or Error. The combination of type and class specifies the content format and interpretation of the message payload. For example, the `retrieve firmware version' type has no parameters as a Command, but contains a string with the firmware version as Reply. Slow-control messages are the primary method for remote control and has functions for moving the software state-machine, request or set process-variables and many others. Messages are bundled together at the \gls{mcf} layer, binding multiple messages into a single payload for efficiency. The lowest slow-control layer is \gls{srp} and is responsible for transmission control. It implements a simple packet-based re-transmission scheme where packet contains an identifier ordinal which must be acknowledged within a $200\,$ms window and is otherwise re-transmitted. The message is resend, with a small delay in between each attempt, to a maximum of 6 times if no acknowledge is received after which it is deemed lost. 

The primary way to set configuration values or get sensor data from remote is by using the process-variable mechanism. Process-variables are defined in a json5-formatted\cite{json5} variable dictionary file and using a code generation tool various files representing the variable dictionary for different programming languages can be generated. For the embedded software C-code is generated for the defined variables, while for remote Java-code is generated. 
 For the embedded software all process-variables are exposed as plain-C variables and are accessible for read and write but from remote access may be restricted during a run, or in the case of sensor data always be read-only.

\section{Build environment}
The build environment consists of four GIT repositories:
\begin{itemize}
	\item \textbf{clb-sw} - Builds program binary for different applications (DOM, DU-Base, CU-Base and golden) and hardware versions (CLB v2 and v4, due to different peripherals). It executes tests and builds documentation.
	\item \textbf{wrpc-sw} - A special branch of the WhiteRabbit PTP core is maintained and the program binaries for the WhiteRabbit PTP core for both the standard and the KM3NeT specific versions is build from this.
	\item \textbf{clb-hdl} - Builds the FPGA image binaries (gateware) for different hardware versions (CLB v2 and v4). This project also uses parts of the WhiteRabbit gateware respository to build the WhiteRabbit PTP core logic.
	\item \textbf{clb} - Includes all of the above projects as GIT sub-modules. It contains the super-build script  which builds all sub-projects and merges all generates binaries into a number of application and hardware specific firmware images.
\end{itemize}
CMake is used as the primary build tool. The entire project can be build inside Gitlab-CI, using a KM3NeT specific docker container.

\section{Conclusions}
The First Generation KM3NeT embedded software has been in development since 2012 and works reliably on the seafloor since 2016 but is now scheduled to be superseded by the Next Generation KM3NeT firmware. The NG firmware has an improved software design for consistent control, contains more reliability features for robust operation and the project now uses modern development tools for reliable and consistent builds. The NG firmware is currently being tested and on track to be deployed fall 2021.

\end{document}